%% file: arxiv.tex
\documentclass[a4paper]{article}

\usepackage{rotating}
\usepackage{booktabs}
\usepackage{layouts}

\usepackage{tikz}
\usetikzlibrary{arrows,mmt}
\usepackage{url}
\usepackage{wrapfig}
\usepackage{amsmath,amssymb}
\usepackage{xspace}
\usepackage{listings}
\usepackage{graphicx}

\usepackage{paralist}
\usepackage{xcolor}

\usepackage{fr-macros/basics}
\usepackage{local}
\usepackage[bookmarksnumbered,bookmarksopen]{hyperref}

\title{Making Isabelle Content Accessible in Knowledge Representation Formats}
\author{Michael Kohlhase\\FAU Erlangen-N\"urnberg\and
Florian Rabe\\FAU Erlangen-N\"urnberg\and
Makarius Wenzel\\Augsburg, Germany}
\begin{document}
\maketitle

%%FIXME draft
%\newpage\tableofcontents\newpage

\begin{abstract}
  
  The libraries of proof assistants like Isabelle, Coq, HOL are notoriously difficult
  to interpret by external tools: de facto, only the prover itself can parse and process
  them adequately. In the case of Isabelle, an export of the library into a FAIR
  (Findable, Accessible, Interoperable, and Reusable) knowledge exchange format was
  already envisioned by the authors in 1999 but had previously proved too difficult.
  
  After substantial improvements of the Isabelle Prover IDE (PIDE) and the \omdoc/\mmt format since then, we
  are now able to deliver such an export.
  Concretely we present an integration of PIDE and \mmt that allows exporting all Isabelle libraries in \omdoc format.
  Our export covers the full Isabelle
  distribution and the Archive of Formal Proofs (AFP) --- more than 12 thousand theories and
  locales resulting in over 65\,GB of OMDoc/XML.  

  Such a systematic export of Isabelle content to a well-defined interchange format like \omdoc enables many
  applications such as dependency management, independent proof checking, or library search.
\end{abstract}

\section{Introduction and Related Work}\label{sec:intro}
\input{intro}

\section{Isabelle and PIDE}\label{sec:isabelle}
\input{isabelle}

\section{OMDoc and MMT}\label{sec:omdoc}
\input{omdoc}

\section{Logical Aspects of the Translation}\label{sec:logical}
\input{logical}

\section{Technical Aspects of the Translation}\label{sec:technical}
\input{export}

\section{Enabled Applications}\label{sec:applications}
\input{applications}

\section{Conclusion and Future Work}\label{sec:conc}
  \input{conc}

% \bibliographystyle{alpha}
% \bibliography{paper,fr-macros/bib/systems,fr-macros/bib/pub_rabe,fr-macros/bib/rabe,fr-macros/bib/institutions}
\newcommand{\etalchar}[1]{$^{#1}$}

\end{document}

%% file: intro.tex
\subparagraph{Motivation}
A critical bottleneck in the field of interactive theorem proving is the lack of interoperability between proof assistants and related tools.
This leads to a duplication of efforts: both formalizations and auxiliary tool support (e.g., for automated proving, library management, user interfaces) cannot be easily shared between systems.
This situation is well-understood by the community and has persisted for decades despite occasional attempts to achieve interoperability by standardization or library translations. 

The story of this article started in 1999, when one author (Kohlhase, who worked on the OMDoc interchange format \cite{omdoc} for formal libraries) wrote an email to another one (Wenzel, who worked on the Isabelle proof assistant \cite{paulson700,LCF-to-Isabelle-HOL:2019}) asking about the status of ongoing efforts to export Isabelle theories in some format that could be further transformed into \omdoc.
Just 19 years later, Wenzel replied to the same email announcing that an Isabelle$\to$OMDoc export now works routinely.
Critically, this export was enabled by the PIDE and \mmt infrastructures developed for Isabelle by Wenzel resp.~for \omdoc by Rabe in the interim.
Despite this massive groundwork laid in the last two decades, the export itself still required about 9 person-months to implement.
This paper tells the story of how we achieved this export after such a long time.

Isabelle99 (October 1999) was a rather small experimental proof assistant for multiple object logics, with $\approx$ 1\,MB source text the for Isabelle/ZF library and $\approx$ 3\,MB for Isabelle/HOL.
The ZF library was particularly interesting for Kohlhase at that time and considered large.
In contrast, Isabelle2020 (April 2020) includes $\approx$ 2\,MB material for ZF and $\approx$ 30\,MB for HOL, or rather $\approx$ 160\,MB if the Archive of Formal Proofs (AFP) is included.
The PIDE/\mmt work flow described in this paper requires a server-class machine to handle all this material: 80\,GB RAM, 8 CPU cores, and 22\,h elapsed time (this includes theory and proof processing by Isabelle).
Thus, a major portion of publicly known Isabelle content\footnote{In the Isabelle community, contributions are usually submitted to AFP for long-term maintenance, and thus become centrally accessible. Only a few exceptional projects are maintained independently (e.g. seL4 \url{https://sel4.systems} or IsaFoR \url{http://cl-informatik.uibk.ac.at/isafor}).} becomes accessible as XML in the \omdoc format: 65\,GB uncompressed or 300\,MB with XZ compression.

\subparagraph{Related Work}
In both formalizations and auxiliary tool support, previous work has shown significant potential for knowledge sharing.
Regarding sharing among proof assistants, library translations such as \cite{hol_isahol,Krauss-Scalable,hol_coq,isahol_isazf} have been used to transport theorems across systems.
An unusual approach is virtualization of HOL4 in Isabelle \cite{Immler-Raedle-Wenzel:2019}, where the ML environment of Isabelle is carefully
instrumented to load the HOL4 library sources (also in ML) and reconstruct theories and proofs within the Isabelle/Pure inference
kernel.
%The virtual HOL4-Isabelle connection works thanks to a coincidence of
%the underlying ML platform (Poly/ML). This it is analogous to the
%coincidence of Scala for Isabelle/PIDE vs.\ MMT/OMDoc, but here we
%merely treat already generated static content and could alternatively
%produce output for a different language (e.g.\ Dedukti).

Most of these approaches produce an isolated image of the source library within the target library.
Alignments \cite{KKMR:alignments:16} have been used to match pragmatically corresponding concepts defined in different libraries \cite{hol_isahol_matching}.
In contrast, \cite{Immler-Raedle-Wenzel:2019} connects interesting results via \emph{lifting and transfer}, where only the
signatures of the main conclusions need to be taken into account.

Regarding sharing among proof assistants and auxiliary tools, Isabelle's Sledgehammer \cite{isahol_fol,LCF-to-Isabelle-HOL:2019} provides a generic way to integrate different automation tools, and Dedukti \cite{dedukti} has been used as an independent proof checker for various proof assistant libraries.
Premise selection tools use, e.g., machine-learning \cite{holyhammer}, to reduce the search space when running automated provers on subgoals.
In all cases, a single tool could be used for every proof assistant --- provided the language and library are available in a universal format that can be plugged into it.
 
Unfortunately, the latter point --- the universal format --- is often prohibitively expensive for many interesting applications.
Firstly, it is extremely difficult to design a universal format that strikes a good trade-off between simplicity and universality. 
And secondly, even in the presence of such a format, it is difficult to implement the export of a library into it.
Here it is important to realize that any export attempt is doomed that uses a custom parser or type checker for the library --- only the internal data structures maintained by the proof assistant are informative enough for most use cases. 
Consequently, only expert developers can perform this step, and of these, each proof assistant community only has very few.

In previous work, the authors have developed such a universal format \cite{omdoc,RK:mmt:10,KR:qed:14} for formal knowledge: \omdoc is an XML language geared towards making formula structure and context dependencies explicit while remaining independent of the underlying logical formalism.
We also built a strong implementation --- the \mmt system --- and a number of generic services, e.g., \cite{rabe:ui:14,mathwebsearch}.
In the DFG-funded OAF Project (Open Archive of Formalization), we have developed export for Mizar \cite{IanKohRabUrb:tmmliotaa13}, HOL Light \cite{KalRab:hollight:14}, IMPS \cite{imps_oaf}, PVS \cite{KMOR:pvs:17}, and Coq in \cite{MRS:coq:19}.
 In what we now call the \emph{OAF approach}, we systematically
\begin{enumerate}[(i)]
\item  defined the logic of the proof assistant in a logical framework by hand, 
\item  instrumented the proof assistant to export its libraries, and 
\item use the instrumented prover to export the libraries
\end{enumerate}
for all these exports. 
\mmt provides the semantics that ties together the three involved levels (logical framework, logic, and library) and provides a uniform high-level API for further processing.
\cite{KohRab:eempal20} gives on overview over the theoretical, technical, and social challenges of the OAF exports. 

In the work reported here, we follow this basic recipe with a few modifications.
Firstly, because Isabelle already includes a logical framework, we do not encode Isabelle in yet another one.
Instead, we extend the existing LF formalization in \mmt to obtain one for the Pure framework of Isabelle.
There are two reasons for this choice: it is conceptually appropriate as it puts the logics defined in Isabelle on the same levels as those defined in other logical frameworks (e.g., \mmt/LF/HOL Light and \mmt/Isabelle/HOL); but it was also necessary for scalability: another layer of logical framework-encoding would make this difficult undertaking even less feasible.
Secondly, Isabelle is extremely complex, and a large portion of our work went to streamlining Isabelle components to enable step (ii) above.
Thirdly, the resulting exports of the Isabelle libraries were significantly larger than any exports we had handled previously.
Therefore, we had to develop new optimizations both on the Isabelle and on the \mmt side to be able to carry out step (iii) above.

%It is critical that the exports systematically avoid any (deep) encoding of logical features.
%That is important so that further processing can work with the exact same structure apparent to a user of the proof assistant. \ednote{Talk about the the rest of the OMDoc/MMT import/export business}

\subparagraph{Contribution and Overview}
We apply our approach to Isabelle \cite{LCF-to-Isabelle-HOL:2019}: we present a definition of the Isabelle logical framework in \mmt and an export feature for Isabelle logics and libraries.
We exemplify the latter by exporting the standard Isabelle distribution \cite{Isabelle:URL} and the \emph{Archive of Formal Proofs} \cite{AFP:URL}.
The translated libraries are available at \url{https://gl.mathhub.info/Isabelle} as compressed OMDoc files.

We present preliminaries about Isabelle and PIDE as well as \omdoc and \mmt in Sections~\ref{sec:isabelle} and~\ref{sec:omdoc}.
Then we describe the logical and the technical aspects of the export in Sections~\ref{sec:logical} and~\ref{sec:technical}.
We sketch some applications enabled by the export in Section~\ref{sec:applications}.

It is difficult to estimate the total workload covered by this paper because it builds on decades of implementation work in both Isabelle and \mmt, much of which was never published in itself.
But concretely for this particular export, we spent about 1 person-month on the overall design of the translation and the implementation, 6 person-months on the implementation on the Isabelle side, 1 on the \mmt side, and 1 on administrative parts and dissemination of the results.

\subparagraph{Acknowledgments}
The authors were supported by DFG grant RA-18723-1 OAF and EU grant Horizon 2020 ERI 676541 OpenDreamKit.

%%% Local Variables:
%%% mode: latex
%%% mode: visual-line
%%% fill-column: 5000
%%% TeX-master: "paper"
%%% End:

%  LocalWords:  omdoc ednote hol_coq,hol_isahol,isahol_isazf hol_isahol_matching isahol_fol Dedukti dedukti mathwebsearch IanKohRabUrb:tmmliotaa13 Logosphere logosphere ObuaImporting Virtualization Immler-Raedle-Wenzel:2019 standardization hol_isahol,Krauss-Scalable,hol_coq,isahol_isazf isahol_fol,isabelle paulson700,LCF-to-Isabelle-HOL:2019 KohRab:eempal20

%% file: isabelle.tex
\subparagraph{The Isabelle Platform}
Isabelle \cite{paulson700,LCF-to-Isabelle-HOL:2019} is a generic platform for formal logic tools.
Its foundation is the \emph{Pure logical framework} by Paulson \cite{paulson700} based on a minimal intuitionistic higher-order logic with declarative natural deduction proofs.
Isabelle/Pure is used to represent object-logics like Isabelle/FOL, Isabelle/ZF, and the most widely used Isabelle/HOL based on Church's simple type theory and Gordon's HOL \cite{Gordon:1985:HOL}.

Extra-logical tools are implemented in the \emph{Meta Language (ML)} in LCF style \cite{Gordon-Milner-Wadsworth:1979}.
Isabelle/ML has full access to the symbolic representation of the logic and provides many add-ons such as concrete syntax and context management for proof tools.
The ML compiler and toplevel environment are managed within the same formal context as the logic, so ML declarations follow the structure of theory specifications and proofs.

ML is mainly used for pure mathematical programming with limited access to the physical world.
Additionally, Scala (running on the Java platform) is used for external tooling: it manages ML processes, formal sources, and the resulting content, and provides an outer shell for Isabelle systems programming with access to GUI frameworks, TCP servers, database engines, etc.
The programming style of Isabelle/Scala resembles Isabelle/ML, and some important modules are available on both sides (e.g.\ formatting of pretty-printed text).

Isabelle's Prover IDE framework PIDE \cite{Wenzel:2014:ITP-PIDE} integrates all development into the semantic text editor Isabelle/jEdit
\cite{isabelle-jedit}.
While the user is composing text, PIDE provides real-time markup about its meaning --  rendered as, e.g., text color, squiggly underline, tooltips, hyperlinks, icons in the border.
The Prover IDE supports ML development as well: users can edit theory sources with embedded ML modules directly, while the ML compiler does static
checking and dynamic evaluation on the spot.
Thus Isabelle has no need for externally compiled modules, in contrast to, e.g., Coq plugins.

More recently, Isabelle/PIDE has been refined to support \emph{headless mode}, which lets a function in Isabelle/Scala observe this markup while a formal library is processed in Isabelle/ML.
Compared to traditional batch-builds, headless PIDE provides more detailed feedback from the prover and more flexibility in dynamic loading and unloading of theories.
In particular, it allows the processing of Isabelle content for other purposes than editing it in a GUI.
This is the central interface that we  use in the work reported in this article.

\subparagraph{Isabelle Libraries} The standard distribution of
Isabelle includes the Isabelle/HOL library with many examples, but the
bulk of applications is in the \emph{Archive of Formal Proofs} (AFP),
which is organized like a scientific online journal.  In April 2020,
AFP had 528 articles by 347 authors, comprising a total of 130\,MB of
source text in 5343 theory files.

Formal processing of the Isabelle distribution plus AFP requires
$\approx$ 46h CPU time or 13h elapsed time, using standard hardware
with 8 CPU cores and 16\,GB RAM. Such \texttt{isabelle build} jobs
\cite{isabelle-system} produce heap images for the internal state of
Isabelle/ML and optional HTML/PDF documents that resemble conventional
mathematical texts.

\subparagraph{Library Structure}
Isabelle libraries consist of formal documents \cite{Wenzel:2019:MKM} structured according to session definitions, theory imports, and commands within theories:
\begin{itemize}
\item A \emph{session} is a collection of theories with optional {\LaTeX} document preparation.
  It may refer to a single \emph{parent session} and multiple \emph{import sessions} (to reuse some of their theories by reloading their sources within the original session name space).
  For example, the session \verb,HOL, is the basis for most applications, and the session \verb,HOL-Analysis, is a substantial library of standard
  mathematics.
  In the AFP, each entry (or ``article'') usually corresponds to a single session with its own setup for the published PDF document.
\item A \emph{theory} is a linear arrangement of commands corresponding to definition--statement--proof in conventional
  mathematical texts.
  The theory header imports multiple parent theories, taking a strictly monotonic \emph{merge} of existing theories as basis for the new one.
  For example, theories like \verb,HOL.Nat,, \verb,HOL.List, are stepping stones towards \verb,Main, and \verb,Complex_Main,, which have global names and are the key entry-points for applications.
\item A \emph{command} is a functional update on the theory context (or proof state) using concrete syntax within the source file.
  Command syntax may embed embed user-defined sublanguages delimited as so-called ``cartouches'', e.g.\ \textbf{ML}~\raise.3ex\hbox{$\scriptscriptstyle\langle$}\emph{val a
        = 1}\raise.3ex\hbox{$\scriptscriptstyle\rangle$}.
  Theories may define new commands at any time --- even Isabelle/Pure itself is defined in user-space relying only on the \textbf{ML} command for bootstrapping.
  For example, the commands \textbf{definition}, \textbf{inductive}, \textbf{fun} define constants and automatically prove characteristic theorems over them.
  The commands \textbf{lemma}, \textbf{proof}, \textbf{qed}, \textbf{by} are used in user-written proofs in the Isar proof language.
\end{itemize}

The overall graph of sessions and theories is managed by Isabelle to exploit parallel processing within multithreaded ML (and Scala).
For example a theory could already be finished on the surface but some of its proofs still pending in parallel forks.
Isabelle/Scala provides operations to explore sources down to command spans (keyword with argument tokens), without requiring a prover process to interpret them in the formal context.

\subparagraph{Library Processing}
The library sources are processed by feeding them to the Isabelle/ML session managed by Isabelle/Scala.
This constructs formal meaning that is a-priori opaque, i.e., a matter of the private context of the logic or user-defined sublanguage.
In order to expose some aspects of the meaning, Isabelle/ML supports several formal message channels:

\begin{itemize}
\item \emph{Output} of regular messages, warnings, errors, etc. with text that typically refers to logical types and terms.
  Pretty-printing with blocks and breaks is supported by default: the front-end usually does the formatting based on precise
  window and font sizes.
  For example, the command \textbf{term} turns its source argument into an internal term and pretty-prints the result with markup to link constants to their definitions.
\item \emph{Reports} to assign markup to existing input sources (with precise positions).
  For example, after reading a term from the source text its precise positions of free and bound variables are reported as XML markup elements \verb,<free/>, and \verb,<bound/>,.
  The editor turns this into the usual Isabelle color scheme of blue vs.\ green variables.
\item \emph{Exports} to attach arbitrary blobs to a theory (with hierarchic names separated by slash).
  For example, the command \textbf{export\_code} turns Isabelle/HOL specifications into program source (for SML, OCaml, Scala), and the result becomes an export artifact of the enclosing theory.
  Thus the current version of input sources (e.g., an open buffer in Isabelle/jEdit) is augmented by the result of \textbf{export\_code} seen as a mathematical function; the editor shows the result via the \emph{virtual file-system} URL \verb,isabelle-export:, within its File Browser, independently of the accidental state of the physical file-system.
\end{itemize}

The exposed aspects of document meaning are stored within the \emph{session database}.
For conventional batch-builds, that is an SQLite database file used like an archive with XZ-compressed entries, and the command-line tool \verb,isabelle export, lists and extracts its content.
For PIDE processing, the database consists of Scala values within the document snapshot and may be explored via user-provided Scala functions, e.g., for GUI painting of annotated document source.
It is also possible to write out the data to another database (e.g., PostgreSQL is supported routinely), or in a completely different application, which is what we do in the OAF-style export reported on in this article. 

To support the latter, Wenzel has modified the processing to allow for application-specific ML functions for presentation.
Whenever a theory node with all its imports is fully consolidated (parallel proofs finished), user-defined ML functions can access its list of commands paired with the internal theory context at each step.

Isabelle/Pure and Isabelle/HOL provide standard presentation functions to expose core material from the logical context, guarded by option \verb,export_theory,.
Results are exported to the session database, using a private XML representation, Isabelle YXML transfer syntax, and XZ compression of the resulting blob.
This works both for batch sessions (\texttt{isabelle build}) and  for headless PIDE sessions (\texttt{isabelle dump}).
Thus, with the current infrastructure, the request by Kohlhase from 1999 could be fulfilled on the spot via \texttt{isabelle dump -B ZF}, but instead of digesting raw XML/YXML data it is better to use typed APIs in Isabelle/Scala (by using module \verb,Export_Theory, as we do in Section~\ref{sec:export}).

%% file: omdoc.tex
\subparagraph{Language}
\omdoc \cite{omdoc} (short for \textbf{O}pen \textbf{M}athema\-ti\-cal \textbf{Doc}uments) is a semantics-oriented XML-based markup format for STEM-related documents.
It conceptualizes mathematical objects in three levels as seen in Figure~\ref{fig:levels}: the \emph{object} level for mathematical formulas and their presentations, the \emph{statement} level for definitions, theorems, proofs, etc, and the \emph{theory} level for collections of statements.
Each level comes in two dimensions for the formal representations of the content addressed to mathematical software systems and the narrative structure addressed to humans.
Higher levels may contain expressions of lower ones, and mixtures of dimensions are allowed, leading to a overall format that can handle flexible levels of formality (see~\cite{Kohlhase:tffm13} for a discussion).

\begin{wrapfigure}r{8cm}\vspace*{-1em}
  \begin{tabular}{|l|l|l|}\hline
    level & formal  & narrative \\\hline\hline
    object & OpenMath & presentation MathML \\\hline
    statement &  sequents & paragraphs + cues\\\hline
    theory & theories/views & sections, etc.\\\hline
  \end{tabular}\vspace*{-.5em}
  \caption{Three level \& two dimensions in OMDoc}\label{fig:levels}\vspace*{-.5em}
\end{wrapfigure}

Even at the early state in 1999, \omdoc already had this general architecture and was therefore well-suited in principle for representing Isabelle content, in particular the Isar proof language \cite{Wenzel:1999:TPHOL} that was new at the time.
But the formal part of \omdoc was purely descriptive and lacked a rigorous semantics.
In particular, the role of the logical systems needed for formally stating mathematical properties was almost fully unspecified beyond the idea --- inherited from OpenMath --- that logics are theories as well. 

Later \mmt (Meta Meta Theories) \cite{RK:mmt:10} re-conceptualized and refined the formal fragment of \omdoc, greatly enhancing both rigor and expressivity.
It models formal objects and statements using logical frameworks, in particular the judgments-as-types paradigm, and bases \omdoc's theory level on the category of theories and theory morphisms following the development graphs approach \cite{devgraphs}.
The former allows for fine-grained specifications of the semantics of individual objects, and the latter allows for inducing and translating knowledge across theories.
A new \emph{meta-theory} relation links a logical framework to the logics defined in it, thus formalizing the ``logics-as-theories'' approach. 

\subparagraph{The \mmt System}
The \omdoc/\mmt language is implemented in the \mmt system (Meta Meta Toolset; see~\cite{rabe:howto:14}), which provides an API for the language constructs at all levels and provides both logical services such as type reconstruction and rewriting and knowledge management services such as IDE and HTML presentation and browsing of libraries.

Because it avoids committing to a specific semantics or logical foundation, foundation-dependent services and features (e.g., type reconstruction) are implemented by splitting the algorithms into a foundation-independent kernel that is user-extensible with foundation-specific rules.
For example, the logical framework LF \cite{lf} is implemented using about 10 rules taking only a few lines of code each.

\subparagraph{Theory Graphs}
Theory graphs are diagrams in the categories of theories and morphisms.
The possible morphisms in \mmt are \textbf{inclusions}, which import all declarations from the domain to the co-domain, \textbf{structures}, which are like includes but copy and translate all declarations, \textbf{views}, which are semantics-preserving translations from domain to codomain, and the \textbf{meta-theory}-relation, which behaves like an include for most purposes.

\begin{wrapfigure}{l}{0.43\textwidth}\vspace*{-.75em}
\providecommand\cn[1]{\ensuremath{\mathsf{#1}}}
\providecommand\myyscale{1}
\providecommand\myxscale{1}
\def\outerthysep{.3mm}
\def\innerthysep{.5mm}
\tikzstyle{thy}=[draw,outer sep=\outerthysep,rounded corners,inner sep=\innerthysep]
\newcommand{\mmtarrowtipmonoright}{right hook}
\newcommand{\mmtarrowtip}{angle 45}
\tikzstyle{meta}=[dotted,-\mmtarrowtip,thick]
\tikzstyle{include}=[\mmtarrowtipmonoright-\mmtarrowtip,thick]
\tikzstyle{view}=[preview,-\mmtarrowtip]
\tikzstyle{struct}=[-\mmtarrowtip,thick]
\usetikzlibrary{decorations,decorations.pathmorphing,decorations.markings}
\tikzstyle{preview}=[decorate, decoration={coil,aspect=0,amplitude=1pt,
                                           segment length=6pt,
                                           pre=lineto,pre length=3pt,
                                           post=lineto,post length=5pt},
                                thick]
\begin{tikzpicture}[xscale=\myxscale,yscale=\myyscale]
% \draw[view] (-4,2.4) --node[above] {translation} +(2,0);
% \draw[struct] (-4,1.7) --node[above] {import} +(2,0);
% \draw[meta] (-4,1) --node[above] {meta-theory} +(2,0);
\node[thy] (lf) at (0,2.5)  {$\cn{LF}$};
\node[thy] (lfx) at (1.8,2.5)  {$\cn{LF+X}$};
\node[thy] (fol) at (-1,1.5)   {$\cn{FOL}$};
\node[thy] (hol) at (.9,1.5) {$\cn{HOL}$};
\node[thy] (mon) at (-2.5,0) {$\cn{Monoid}$};
\node[thy] (gp) at (-.5,0) {$\cn{CGroup}$};
\node[thy] (rg) at (2,0)  {$\cn{Ring}$};
\node[thy] (zfc) at (-2.8,1) {$\cn{ZFC}$};
\draw[meta](lf) -- (fol);
\draw[meta](lf) -- (hol);
\draw[meta](fol) -- (mon);
\draw[meta](fol) -- (gp);
\draw[meta](hol) -- (rg);
\draw[include](lf) -- (lfx);
\draw[view](fol) -- node[above] {\footnotesize$\cn{f2h}$} (hol); %$
\draw[struct](gp) to[bend right=10] node[above]
{\footnotesize$\cn{add}$} (rg); %$
\draw[struct](mon) to[out=20,in=160] node[above]
{\footnotesize$\cn{mult}$} (rg); %$
\draw[include](mon) -- (gp);
\draw[meta] (fol) -- (zfc); %$
\draw[view] (fol)  to[bend right=30] node[above,near start]{\footnotesize$\cn{folsem}$} (zfc); %$
\draw[view] (mon) -- node[left]{\footnotesize$\cn{mod}$} (zfc); %$
\end{tikzpicture}
\vspace*{-.2em}
\caption{Meta-Levels in OMDoc/MMT}\label{fig:tgraph}
\vspace*{-1em}
\end{wrapfigure}

Figure~\ref{fig:tgraph} shows an example of a typical setup of formalizations in \mmt: Dotted lines represent the meta-theory-relation, hooked arrows are includes, squiggly arrows represent views, and the normal arrows represent named structures.
Here LF is used as a logical framework to define some logics, which are then used as meta-theories for algebraic theories.
We see three pragmatic levels: the logical frameworks at the top, logics in the middle, and the domain theories at the bottom.
Meaning trickles down from the theories at the top (the ones without meta-theories), which are implemented directly in \mmt/Scala as described for LF above.
This setup can even encode model theory theory morphisms into semantic theories like ZFC set theory.

%%% Local Variables:
%%% mode: latex
%%% mode: visual-line
%%% fill-column: 5000
%%% TeX-master: "paper"
%%% End:

%  LocalWords:  omdoc omdoc athema uments conceptualizes sequents re-conceptualized devgraphs formalizing mmtsys lf myyscale myxscale outerthysep innerthysep tikzstyle draw,outer outerthysep,rounded corners,inner newcommand mmtarrowtipmonoright mmtarrowtip mmtarrowtip,thick usetikzlibrary decorations,decorations.pathmorphing,decorations.markings coil,aspect 0,amplitude lineto,pre lineto,post fig:tgraph Kohlhase:tffm13 0,amplitude

%% file: logical.tex
The logical basis of our export is a definition of Pure in the \mmt system.
\mmt allows defining a wide variety of logical frameworks, and we use PLF as a starting point, a polymorphic variant of LF \cite{lf} that already exists in the \mmt standard library \cite{MR:prototyping:19}.

\subsection{Type System and Logic}

\subparagraph{Types, Terms, Propositions}
We use a shallow embedding of Pure in PLF.
Besides simplicity, this has a critical scalability advantage: a deep embedding would lead to substantially larger PLF-expressions when already our shallow embedding ended up yielding the largest export size we had ever attempted (since then eclipsed only by our analogous export for Coq \cite{MRS:coq:19}).
Consequently, as Pure uses shallow polymorphism (type variables bound at the outside of declarations), we cannot use LF itself but need to extend it with shallow polymorphism.
That is why we use PLF instead.

Using a shallow embedding, most Pure primitives are represented as their PLF-counterparts:
Pure-types and terms are represented as PLF-types and terms.
This includes in particular Pure's simple \textbf{function} types, $\lambda$-abstractions, and application.

The remaining primitives can simply be declared as PLF-constants.
That yields a PLF-theory containing in particular the constants
\begin{compactitem}
\item $\prop:\type$ for the type of \textbf{propositions},
\item $\ded:\prop\to\type$ mapping each proposition $\phi$ to the type $\ded\, \phi$ of \textbf{proofs} of $\phi$.
\end{compactitem}

\noindent That is the bare minimum to connect Isabelle/Pure to PLF: the remaining connectives are
produced from the regular export of the \verb,Pure, theory itself, yielding further constants:
\begin{compactitem}
\item $\textit{Pure.eq}:\tPi[\type]{a}a\to a\to\prop$ polymorphic \textbf{equality} (with implicit $\alpha\beta\eta$-conversion),
\item $\textit{Pure.all}:\tPi[\type]{a}(a\to\prop)\to\prop$ for the polymorphic binder for \textbf{local parameters},
\item $\textit{Pure.imp}:\prop\to\prop\to\prop$ for the constructor for \textbf{logical entailment}.
\end{compactitem}
Relative to these declarations, it is straightforward to translate all Isabelle types, terms, and propositions.

\subparagraph{Proof Terms}
Like LF but unlike Pure, PLF offers dependent types.
These are not needed for representing the simply-typed Pure language but are helpful to concisely represent Pure-proofs as PLF-terms in Curry-Howard style.
Thus, Pure proof terms can be exported analogously to types, terms, and propositions.
However, in practice, we only export proof terms for small examples because proof terms for actual Isabelle/HOL are far too big.
After our work on Isabelle, we conducted a similar export for Coq in \cite{MRS:coq:19}.
Here we included proof terms, and the sizes, while large, remained manageable.
But due to the lack of Coq-style implicit computation, we expect Pure proof terms to be even larger.

However, there is a separate, deeper reason to defer proof exports: it is still unclear what the best way to export proofs is.
The export of low-level proof terms is straightforward, but the proof objects are huge and have only limited value (independent proof checking being the main one).
The high-level proofs seen by the user are much more interesting but lack the information inferred by the prover.

Therefore, we opted for exporting all proofs as dummy terms that carry only the information that the theorem was checked by Isabelle and which dependencies were used.
Additionally, we include, as an informal narrative text, the command-source of the Isar text: this treats the whole proof as one unit, without the hierarchical structure of Isar proofs (see also the discussion in~\ref{sec:ontology} and \ref{sec:cross} below).

\subsection{Declarations}\label{sec:declarations}

\subparagraph{Foundational Declarations}
It is straightforward to represent the foundational declarations of Pure theories as PLF-declarations as follows:
\begin{compactitem}
 \item Pure-\textbf{type operators} $a$ of arity $n$ as $n$-ary PLF-constants
   \[a:\type\to\ldots\to\type\to\type\]
 \item Polymorphic Pure-\textbf{terms} $c$ of type $A$ using type variables $a_1,\ldots,a_n$ as PLF-constants
   \[c:\tPi[\type]{a_1}\ldots\tPi[\type]{a_n}A\]
 \item Polymorphic Pure-\textbf{axioms} $s$ with type parameters $a_1,\ldots,a_n$ asserting proposition $F$ as PLF-constants
   \[s:\tPi[\type]{a_1}\ldots\tPi[\type]{a_n}\ded\,F\]
\end{compactitem}

All three kinds of declarations may carry definitions, which can be represented by giving the PLF-constant a definiens.
This is used only for type operators and term \emph{abbreviations}.
HOL type definitions are a special case of high-level declarations as described below, and Pure turns term definitions are mapped to definition-less constants with defining axioms (multiple ones in case of overloading).
Additionally, theorems are represented using the proof as the definiens (as described above).

\subparagraph{Identifiers}
Isabelle assigns to each foundational declaration a unique identifier.
It uses separate namespaces for types, terms, and theorems and usually qualifies their names by the base name of the enclosing theory.
Every theory exists within an Isabelle session, whose name usually qualifies the theory's base name. Both qualification schemes are optional --- there is no strict enforcement.

For reusability, it is preferable to use a single namespace (to ensure globally unique identifiers for all declarations) and to use a uniform naming schema for all identifiers.
Moreover, \mmt requires all names to be globally unique by qualifying them with an ownership-defining URI.

Therefore, we have chosen the following naming scheme for all declarations:

\begin{quote}
\texttt{https://isabelle.in.tum.de?}\emph{long-theory-name}\texttt{?}\emph{entity-name}\texttt{|}\emph{entity-kind}
\end{quote}
where \emph{long-theory-name} is the session-qualified theory name, \emph{entity-name} the declaration name within the theory context, and \emph{entity-kind} its name space: notably \verb|type|, \verb|const|, \verb|thm|, or other name spaces of user-defined concepts.
For example,
\begin{center}
\texttt{https://isabelle.in.tum.de?HOL.Nat?Nat.nat|type}
\end{center}
refers to the type \texttt{nat} of natural numbers in the theory \texttt{Nat} in the session \texttt{HOL} of the main Isabelle library.
The seemingly redundant repetition of \texttt{Nat} is needed to cover corner cases, including some unqualified names in Isabelle/Pure.

\subparagraph{High-Level Declarations}
Isabelle provides a user-extensible set of high-level specification elements, whose semantics is defined by elaboration into foundational ones.
Examples include HOL-type definitions or the definition of inductive data types and recursive functions.
Similarly, the high-level specification contexts of locales and type-classes (see below) are elaborated into primitive concepts of the logic.
Both are already covered by exporting their elaboration, but that results in representations without the high-level structure seen by users.

\mmt provides a similar extensible declaration pattern mechanism \cite{HorKohRab:emfsl12,MueRabRot:rslffml20} so that we can use them to represent Isabelle's high-level declarations in a structure-preserving way.
We have so far carried out this effort only for locales and leave other elements to future work:
it could be done by a generic Isabelle/ML interface for such specification elements such that the export works uniformly for all its instances.
Then a manageable separate implementation effort would be needed for each specification element.
However, because the individual specification elements were implemented by different authors and can be very complex, no single person could retrofit them to implement this interface, and a long-term community effort is required.

\subsection{Module System}

\subparagraph{Theories}
The \mmt module system subsumes the expressivity of Isabelle theories and is available for every language defined in \mmt such as PLF.
Thus, all Isabelle theories (including those for logics like HOL) are represented straightforwardly as PLF-theories.

\subparagraph{Locales} As the Isabelle logical framework lacks primitive support for ``little theories'', a locale definition is elaborated into a constant definition (predicate) for the logical specification, together with extra-logical management of the resulting context and conclusions produced within it \cite{isabelle_locales}; similar techniques are used for Isabelle type classes \cite{isabelle_classes} on top of locales.

Without any special care, the export of locales merely shows these predicate definitions with theorems depending on additional parameters and premises. But this low-level elaboration is not what Isabelle users users expect.  Instead we refer to exported information about the original structure of locale specifications and map that to first-class theories in \mmt. Subsequently, we illustrate this approach by a representative example.

\subparagraph{Semigroups}
Consider the following locale for semigroups.
It declares (fixes) the binary operation (where we write \verb|x*y| for \verb|op x y|), assumes the associativity axiom, defines the squaring function, and states a simple theorem:
\begin{lstlisting}
locale sg =
  fixes op :: 'a -> 'a -> 'a  (infixl * 70)
  assumes assoc: $\forall$ x y z. (x * y) * z = x * (y * z)
begin
  definition sq :: 'a -> 'a where sq x = x * x
  theorem sqsq: sq (sq x) = x * sq x * x  <proof>
end
\end{lstlisting}
Note that the universe of the semigroup is not declared explicitly.
Instead, Isabelle locales treat any type variable that remains uninstantiated after type-checking as a type fixed in the locale.
In our PLF representation, this convention is made explicit by declaring the universe $a$ as a type and then treating all fixed types and operations uniformly.
In the sequel, we use the words \emph{structure} to refer to a tuple of values interpreting the fixed types and operations, and \emph{instance} for a structure that satisfies the assumed axioms.

\subparagraph{Translation by Elaboration}
The locale's elaboration is represented as the following set of PLF-constants (where we again write $x*y$ for $op\,x\,y$ but note that $op$ is now always a bound variable):
\begin{compactitem}
 \item one membership predicate that ranges over structures and a defining axiom for it that makes it true for instances:
  \[sg: \tPi[\type]{a}\tPi[a\to a\to a]{op}\prop\]
  \[\textit{sg\_def}: \tPi[\type]{a}\tPi[a\to a\to a]{op}\ded\,(sg\,a\,op) \Leftrightarrow \forall x,y,z.(x*y)*z=x*(y*z)\]
 \item for every definition, a global constant and a defining axiom for it, both abstracting over structures:
\[sg.sq:\tPi[\type]{a}\tPi[a\to a\to a]{op}a\to a\]
\[\textit{sg.sg\_def}:\tPi[\type]{a}\tPi[a\to a\to a]{op}\ded\,d=\tlam[a]{x}x*x\]
 \item for every theorem, a global theorem abstracting over structures and relativized to instances:
\[\mathll{sg.sqsq:\tPi[\type]{a}\tPi[a\to a\to a]{op}\ded\,(sg\,a\,op) \impl \forall x. SQ\,(SQ\,x)=x*(SQ\,x)*x\\
   \tb\mmtdef \text{(proof omitted)}}\]
(abbreviating $sq.sq\,a\,op$ as $SQ$).
\end{compactitem}

Note that Isabelle's elaboration introduces the function $sg.sq$ for all structures even though it is only defined for instances.
This is sound in the special case of Isabelle because function types are simple and all types are non-empty (which makes adding unspecified operations conservative) and because all locale theorems are relativized to instances.

\subparagraph{Reconstruction of Isabelle Locales s MMT Theories}
By elaborating locales into global declarations, some information about the modular structure is lost.
To allow for preserving that structure, we additionally and redundantly export every locale as a PLF-theory with the following local declarations:
\begin{compactitem}
 \item a primitive constant for all fixed types and operations and assumed axioms:
  \[a: \type\]
  \[op: a\to a\to a\]
  \[assoc: \ded\,\forall x,y,z.(x*y)*z=x*(y*z)\]
  (writing $x*y$ for $op\,x\,y$),
 \item a defined constant for each definition and theorem:
  \[sq:a\to a  \mmtdef \tlam[a]{x}x*x\]
  \[sqsq:\ded\,\forall x. sq\,(sq\,x)=x*(sq\,x)*x \mmtdef \text{[proof omitted]}\]
\end{compactitem}
This nicely conforms to the intention of Isabelle locales as extra-logical add-ons to the Pure logic.
We represent sublocale relations and locale interpretations as PLF theory morphisms accordingly (by re-using exported information from Isabelle locale management).

\subparagraph{Type Classes}
Type classes are a special case of locales with some add-on infrastructure, notably for type inference.
A locale may become a type class if it has exactly one free type variable \verb,'a,.

If $sg$ is instead declared as a type class, the following additional declarations are present:
\begin{compactitem}
 \item for every fixed operation, a global constant abstracting only over the single fixed type:
  \[sg\_class.op:\tPi[\type]{a}a\to a\to a\]
 \item for every assumed axiom, a corresponding global axiom relativized by the membership predicate $sg$ of the locale (instantiating the fixed operation $op$ with $sg\_class.op\,a$):
  \[sg\_class.assoc:\tPi[\type]{a}\ded\,sg\,a\,(sg\_class.op\,a) \impl \forall x,y,z.(x*y)*z=x*(y*z)\]
(writing $x*y$ for $sg\_class.op\,a\,x\,y$)
 \item for every definition, a corresponding global constant with a defining axiom,
 \item for every theorem, a corresponding global theorem.
\end{compactitem}

\subsection{Ontology}\label{sec:ontology}

The description above covers the translation of all logical content.
But it is useful to additionally export a high-level abstraction of the library ontology in semantic web style.
This includes all named entities (locales, theorems, etc.) and their interrelations but excludes all complex objects (types, terms, proofs).

Such an ontology export is easier to maintain efficiently, e.g., using RDF triple stores.
And it is sufficient for many important applications such as querying the dependency relation between declarations.
Additionally, it can easily include metadata such as check times.

Isabelle/MMT performs such an RDF/XML export as well, see also \ref{sec:amount} for the amount of relational information.
We originally presented this RDF export in \cite{CKMRSW:ulo:19} together with an Upper Library Ontology (ULO) that describes and provides a uniform vocabulary of classes and relations for all proof assistants; therefore, we mention only a few recent improvements here.
The relational ontology also captures some aspects of inductive and primitive recursive definitions (via the binary relation \verb,ulo:inductive-on,).
Most importantly, our export now fully covers dependencies, spanning a large dependency graph over the source text: it relates via the binary relation \verb,ulo:uses, every theorem statement with every used constant and every proofs with every used theorem.

%%% Local Variables:
%%% mode: latex
%%% mode: visual-line
%%% fill-column: 5000
%%% TeX-master: "paper"
%%% End:

%  LocalWords:  lf compactitem ded noindent arity texttt thm sg x,y,z sg.sq tlam mathll
%  LocalWords:  isabelle_classes,isabelle_locales sg.sqsq impl mmtdef sq.sq assoc sq:a sq
%  LocalWords:  sqsq sq sq verb,ulo:inductive-on verb,ulo:uses

%% file: export.tex
The majority of the export is not \omdoc-specific and carried out on the Isabelle side; this appeared first in the official release Isabelle2019 (June 2019), but
the present paper uses the reworked and simplified version of Isabelle2020 (April 2020).
Being integrated into Isabelle has the advantage that most of our work can be immediately reused for exports into other formats than \omdoc.
Only little \omdoc-specific code is necessary for building and serializing the XML objects in \omdoc format.
For this, we use the \mmt API for \omdoc, which is also written in Scala and therefore directly callable from PIDE.
This code is now part of the \mmt distribution (first in release 14 from November 2018).

The resulting inter-dependency between the code bases is handled as
follows: if the \mmt directory is registered to Isabelle as
\emph{component}, it provides a tool \verb,isabelle mmt_build, (shell
script) to build MMT with Isabelle support enabled. The resulting
\verb,mmt.jar, will provide further tools \verb,isabelle mmt_import,
and \verb,isabelle mmt_server, (in Scala) to perform the import and
view its results. Users merely need to invoke, e.g.,
\verb,isabelle mmt_import -B ZF,.

\subsection{Export from Isabelle}\label{sec:export}

Isabelle/Scala provides a standard module \verb,Export_Theory, to expose theory content to other tools via a statically typed API that imitates Isabelle/ML datatypes for types and terms. The communication between Isabelle/ML and Isabelle/Scala works via untyped XML trees, without any special tricks about meta-programming. Instead, sources in both languages reside next to each other in the official Isabelle repository, are manually updated accordingly.

A first version of the Isabelle export facility appeared in Isabelle2018 (August 2018). It was originally motivated by early versions of Isabelle/MMT, and has grown into an independent Isabelle service. It is supported by command-line tools like \texttt{isabelle export} and \texttt{isabelle dump} \cite{isabelle-system}; \texttt{isabelle build} with option \verb,export_theory, exposes logical content as follows.
\begin{itemize}

\item Foundational theory content of the Isabelle/Pure \emph{logical
    framework}: \textbf{types} (base types and type constructors),
  term \textbf{constants} (including functions, binders, quantifiers
  as higher-order constants), \textbf{axioms} (including equational
  axioms that count as primitive definitions), and \textbf{theorems}
  (propositions with a proof).
  Actual proofs are not exported by default --- they are prohibitively
  large. The option \verb,export_standard_proofs, provides proof terms
  in a standardized format that facilitates import in other tools, but
  this only works for small examples so far.

\item Constant definitions of Isabelle/Pure, as a relation between a
  single constant with multiple axioms. Overloading in Isabelle means
  that a polymorphic entity is characterized on multiple
  (non-overlapping) type instances. The majority of constants are
  non-overloaded, with exactly one equational axiom to express its
  definition. This relation of constants to their defining axioms is
  exported, too.

\item Type definitions of Isabelle/HOL in the sense of Gordon and
  Pitts \cite{pitts93}. This axiomatization scheme can be interpreted
  definitionally within the standard semantics of the HOL
  logic. Isabelle/HOL provides a separate module to create new types
  via that mechanism. Some key information is exported: the old
  representing type, the new abstract type, the name of the morphisms
  between the two with the axiom stating the relation.

  This allows recovering HOL typedefs faithfully, where Pure theory
  content would only show the individual particles. It also serves as
  an example to ``query'' derived specification mechanisms in
  Isabelle/ML, to expose its own level of abstraction to the exporter.
  
\item Term constants with indication of derived specifications
  mechanisms, e.g.\ \textbf{primrec} functions, \textbf{inductive} or
  \textbf{coinductive} relations. This works by querying generic
  information in Isabelle/Pure about functional or relation
  specifications (aka. ``Spec Rules''). The Isabelle/HOL
  implementations provide this data on their own account.

  This merely provides a rough classification of term constants at a
  very abstract level. The full complexity of Isabelle/HOL
  specification mechanisms is more difficult to capture: it would mean
  to follow many implementation details, including ones that have
  changed fundamentally over the years of ongoing Isabelle
  development.
\item Dependencies of proven theorems wrt.\ types, consts, theorems,
  as recorded by the Isabelle inference kernel: This spans a large
  dependency graph over the document in terms of the primitive logic
  --- extra-logical aspects are missing (e.g., dependency on
  notation). Partial support for these \emph{proof constants} had been
  part of the Isabelle codebase over many years, but we had to
  rework this substantially to make it suitable for our application.

\item Locales in the sense of Ballarin \cite{Ballarin2014} and type
  classes as special locale interpretations in the sense of Haftmann
  and Wenzel \cite{isabelle_classes,Haftmann-Wenzel:2009}:
%  Isabelle locales are \emph{named contexts} with type parameters
%  (implicit), term parameters (\textbf{fixes}), and premises
%  (\textbf{assumes}). Within such a ``little theory'' it is possible
%  to spell out definitions, statements, proofs as usual --- all
%  results are understood as relative to the context. The foundations
%  will contain an extra prefix of type variables, term abstractions
%  and assumptions according to the context.
  The export of locales preserves some of its internal structure,
  notably the locale dependency relation stemming from the
  construction of locales and sub-locales (by definition), as well as
  later locale interpretations (by proof).
  These are then exported as \mmt theory morphisms.
%  Type classes are special locales with a single type parameter and a
%  canonical locale interpretation to connect \emph{type class
%    parameters} (polymorphic constants with class constraints) to
%  \emph{locale parameters} (fixed variables of the context).
  For type classes, the export shows the canonical locale interpretation but without an
  explicit connection to the type class.
  This would have to be a type-indexed family of \mmt theory morphisms.
\item The order-sorted algebra of \emph{type classes} (subclass
  relation) and \emph{type arities} (image behavior of type
  constructors wrt.\ type class domains and ranges) in the sense of
  \cite{Nipkow-Prehofer:1993}:
  This allows reconstructing Isabelle's built-in type class
  reasoning by an external program (for example, an application could
  give it to a separate process running Isabelle/Pure and reuse the
  original implementation in module \verb,Sorts, Isabelle/ML). An
  alternative is to imitate these operations in a different
  programming language.\footnote{Isabelle/Scala does not provide any
    type-class reasoning on its own, because it is meant to be for
    external system management only. Logical operations are done
    properly in Isabelle/ML.}
\end{itemize}

Formal entities have two name components: \emph{kind} (to distinguish the namespace) and \emph{full name} (usually with the theory base name as qualifier).
In addition, there is an \emph{external name} for printing (partially qualified according to standard namespace policies), a \emph{source position},
and a \emph{command span identifier}.
The latter allows in particular arranging the content according to the order in which it occurs in the source text so that exported types, constants, theorems appear as a digest for each specification element in the text (e.g.\ for \textbf{definition}).

Moreover, if the target format of the export supports references to the original source, this can be used to attach such a reference or even the entire source fragment to each formal entity.
We do that for our \omdoc export.

\subsection{Import into \mmt}\label{sec:import}

\begin{sidewaysfigure}[hp]
  \centering
  \includegraphics[width=\columnwidth]{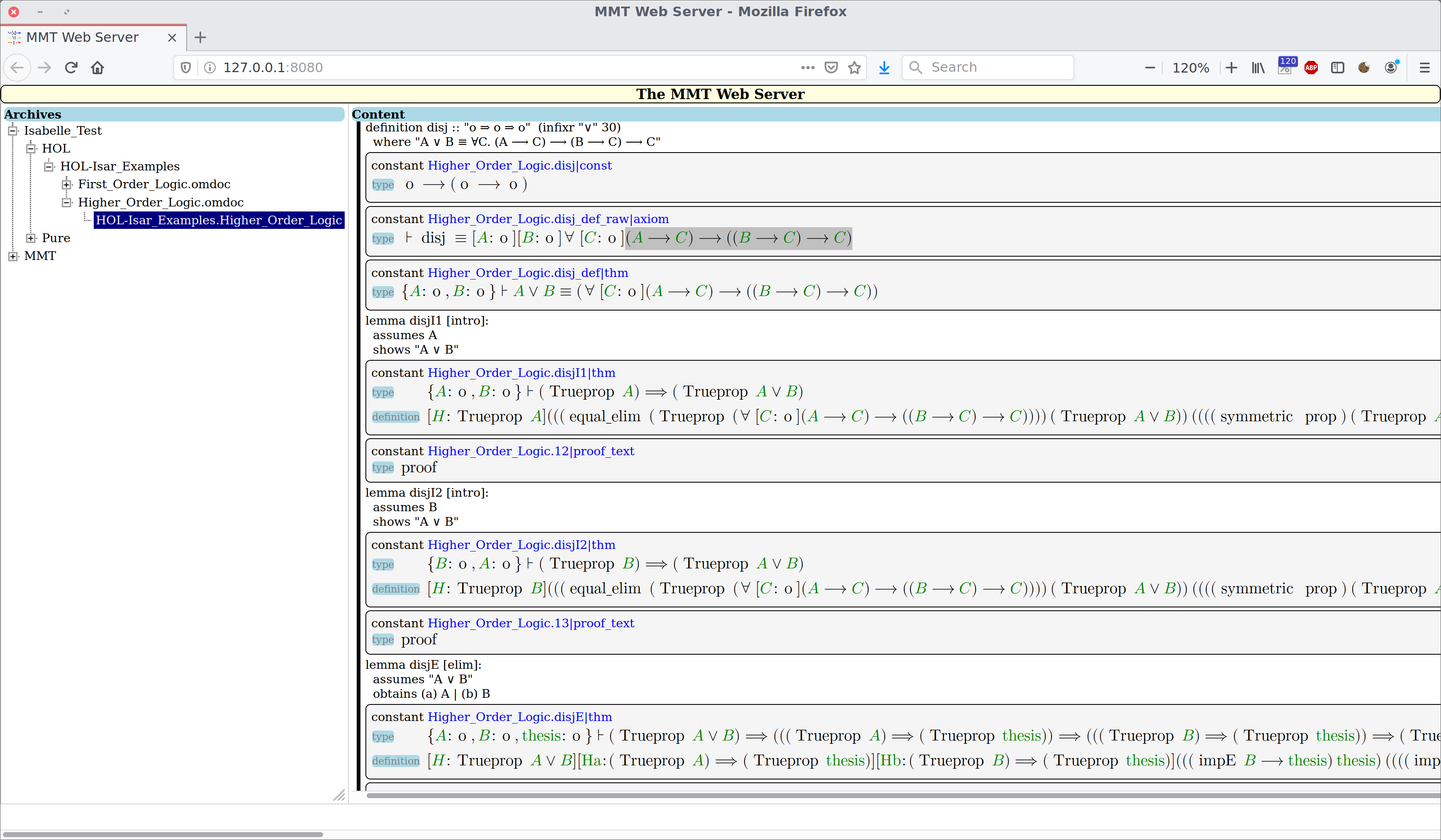}
  \caption{Disjunction in Higher-Order Logic: definitions, theorems and proof terms}
  \label{fig:HOL-disj}
\end{sidewaysfigure}

The entities listed in Section~\ref{sec:export} can be serialized almost directly as \mmt constants relative to the PLF framework as described in Section~\ref{sec:logical}.
That is not surprising as much of that work motivated by the present export in the first place.
Figure~\ref{fig:HOL-disj} shows the \mmt browser displaying an example that is very small and thus includes proof terms.
Note how every formal declaration is preceded with an informal narrative fragment containing the original source text, this is for the orientation for Isabelle users. 

In the sequel, we describe a few specific adaptations of the term language that were required to reconcile traditional Isabelle/ML representations with the more conventional $\lambda$-calculus of PLF in \mmt.

\subparagraph{Type arguments for constants}
The traditional representation of polymorphic constants in Isabelle and the HOL family \cite{pitts93} is to give the full \emph{type instance} at each occurrence in a term, instead of the \emph{type arguments} that produce the instantiation of the general type schema.
For example, constant \verb,id :: 'a => 'a, occurs in particular terms as the pair \verb|(id,| $\tau$ \verb,=>, $\tau$\verb,), for the respective type $\tau$.
This is both redundant (because the type instances are usually bigger than the type arguments) and inconvenient (because it is more difficult to obtain the type arguments from the instantiations than the other way around).
In contrast, PLF treats \verb,id, as a function with dependent type $\tPi[\type]{a} a \to a$ and occurrences are just applications $(\mathit{id} \; \tau)$.

Isabelle/ML provides operations to switch between the two representations within a given context of constant declarations.
Our theory export always uses the second form with type arguments: this reduces the size of exported material and allows importing terms into PLF without again referring to the environment of constant declarations.

\subparagraph{Variable names}
Isabelle variables come in various flavors: free variables (e.g., \verb,x,), schematic variables with index (e.g., \verb,?x10,), and bound
variables (e.g., \verb,x, in $\lambda$\verb,x::,$\tau$\verb,. x,) which
is notation for the de-Bruijn index abstraction
\verb;Abs (x, ;$\tau$\verb;, B.0); where \verb,x, is retained as a
comment).

To fit smoothly into the $\lambda$-calculus of PLF, schematic variables are renamed to fresh free variables.
Since schematic variables are morally like a universal quantifier prefix, this preserves the logical meaning of a statement.
And bound variable comments in abstractions are renamed locally to avoid clashes with free variables in the same scope.
Thus the comment can be used literally in PLF as a named abstraction ignoring the unnamed de-Bruijn index representation of Isabelle.

\subparagraph{Type class constraints}
Isabelle type variables are decorated with type class constraints, e.g., \verb,'a::order, for types that belong to the class \verb,order,
defined in the Isabelle/HOL library (e.g., \verb,nat, with its standard order):
this links certain operations to overloaded term constants (e.g., \verb,less :: 'a => 'a => bool,) and ensures logical premises on these operations (e.g., stating that \verb,less, is a strict order on the type).

Isabelle type class operations are managed by extra-logical means to eliminate the implicit overloading.
In PLF this merely results in multiple constant definitions for different type arguments.

Isabelle class premises become logical constraints in a straight-forward manner: a type class is a predicate over types in PLF.
So \verb,'a::c, means that the predicate \verb,c, applied to type \verb,'a, holds.
Statements with class constraints $\phi($\verb,'a::c,$)$ are augmented by a prefix of preconditions \verb,'a::c,${} \Longrightarrow \phi($\verb,'a,$)$, effectively eliminating the constraint within the logic.

\subsection{Statistics for Isabelle/AFP}\label{sec:amount}

Our test hardware for the \mmt export of Isabelle/AFP is a server machine with 40 CPU cores (80 hardware threads), 128\,GB RAM (2 NUMA nodes), and fast SSD storage.  In the sequel we give an overview of the material for Isabelle2020 (April 2020) with MMT/52adb5e338811e\footnote{\url{https://files.sketis.net/Isabelle_MMT-20200421}}, together with AFP/91f1cdbeefc0 \cite{AFP:URL}: These sources consist of 680 sessions distributed over 7,027 files comprising 160\,MB of theory text (30\,MB XZ-compressed).
The exported content comprises
\begin{compactitem}
\item 7,027 theories and 5,291 locales (``little theories''), including 1,236 type classes,
\item 2,116,638 individuals (11,724 \verb,type,, 204,404
  \verb,const,, 236,186 \verb,axiom,, 1,497,689 \verb,thm,).
\item 400,996,957 relations, including 386,325,246 \verb,ulo:uses,
  (i.e.\ the overall dependency graph of \verb,type,, \verb,const, and
  \verb,thm, items)
\item 65\,GB OMDoc/XML (310\,MB XZ-compressed)\footnote{\url{https://gl.mathhub.info/Isabelle/Distribution/commit/db1009a326c8}
  and \url{https://gl.mathhub.info/Isabelle/AFP/commit/346f28873c9f}}
\end{compactitem}
The entire process of Isabelle/PIDE document checking, export to \mmt, and serialization as XZ-compressed XML requires 80\,GB RAM, 8 CPU cores, and 22h30 elapsed time.
Thus, compared to an elementary batch-build, our export requires around 2 times the memory and 2--5 times the elapsed time (mainly because Isabelle/\mmt uses less parallelization than \texttt{isabelle build}).
We emphasize that these resource figures are for the \emph{entire} AFP, including the special sessions tagged as \verb,slow, or \verb,large, which are often omitted because they take a lot of resources to process.

%These numbers are already the result of several rounds of performance tuning to make the existing PIDE infrastructure scale from %small editor sessions towards large theory libraries.

%That corresponds to the amount of material that is produced from a single Isabelle/ML process with full PIDE markup and export artefacts handed over to Isabelle/Scala.
%In comparison, the bottom-line of resource requirements is given by a regular \texttt{Isabelle build} without the overhead of PIDE document
%processing and exports: 16\,GB RAM, 8 CPU cores, 46h CPU
%time, 13h elapsed time.
%Using 4 parallel processes and 40\,GB RAM, elapsed time shrinks to 4h.
%
%Consequently, there are two main strategies for further scaling of
%Isabelle/MMT export:
%\begin{enumerate}
%\item improving PIDE session management to require fewer resources
%  and/or work with separate parallel processes (instead of just one
%  multi-threaded ML process),
%\item improving batch-builds to export more details, without the
%  full PIDE overhead.
%\end{enumerate}
%  
%It is ultimately a question how much document content we want to see
%in the result: OMDoc intends to cover a broad spectrum of formal and
%informal material, including as much markup as feasible. Other
%applications could restrict themselves to core theory content (types,
%consts, theorems), and thus work with non-PIDE batch-builds and fewer
%resources. But note that proof terms would again increase the order of
%the scalability problem.

The size of the exported OMDoc data structures is linear in the size of the original sources, increased by about factor $10$ in XZ-compressed form.
This increase in size is a gain, not a deficiency --- it stems from the fact that the exported XML contains substantial additional information that is implicit in the sources but extremely difficult to infer:
all occurrences of symbols are disambiguated and exported with their unique URIs;
the exported XML elements carry source references, i.e., URIs that link to the corresponding location in the source;
all type arguments of occurrences of polymorphic constants and all types of bound variables are included in the XML even if omitted in the sources;
and all theorems automatically generated by Isabelle are included in the export.
We could suppress some of this information, but that would defeat the purpose of our export: only Isabelle can infer all details, and handing it to other tools is our export's main value.
The uncompressed XML files are much larger because they are very verbose and optimized for context-free processing.
But we never write the XML directly to the file-system: all reading and writing of XML is filtered through XZ compression.

\subsection{Maintainability}

When developing proof assistant library exports, the challenge of maintainability is often overlooked or underestimated.
This is partly caused by the incentives of the academic system that rewards quickly published results rather than long-term sustainable ones.
We have consciously taken several steps to ensure maintainability.

Firstly, we use statically-typed Scala APIs as much as possible, both in the export from Isabelle and in the import into \mmt.
Almost all the new code we wrote for the occasion was immediately integrated with the existing abstract interfaces.
The remaining glue code that connects Isabelle's abstract export with \mmt's abstract import comprises only a few thousand straightforward lines of code.

Secondly, wherever possible we wrote new code in the Isabelle repository rather than the \mmt repository.
This forces future Isabelle development to maintain our abstract code, in particular when PIDE data structures change.
Concretely, we pushed only the parts of the code that actually depend on the \mmt data structures to the \mmt repository.
That portion consists of only about $2000$ lines of code, mostly straightforward code for creating instances of the \mmt data structures.
The rest of the export code is generally reusable for other Isabelle exports and pushed to the Isabelle repository and already released as an official Isabelle feature.
In fact, this design has already proved beneficial as Wenzel was able to reuse the Isabelle part of our code in a recent export to Dedukti (still unpublished).

Finally, the fact that Isabelle and \mmt can communicate via the Java VM has proved a huge advantage for maintainability.
We were able to design the code in such a way that \mmt is an optional plugin component for Isabelle and vice versa.
Thus, users running Isabelle can simply register \mmt as a plugin with Isabelle and then run \verb,isabelle mmt_import, on the command-line.

Whenever a new Isabelle release is published, it will be a matter to update some statically-typed Scala functions for Isabelle/MMT.
Informed by our experience of multiple similar exports, we judge this one to be the most maintainable export of a proof assistant library so far, in fact by a wide margin.

%\begin{sidewaysfigure}[ph]
%  \centering
%  \includegraphics[width=\columnwidth]{SG-locale.png}
%  \caption{Translation of locale definition and definition in locale context.}
%  \label{fig:SG-locale}
%\end{sidewaysfigure}

%%% Local Variables:
%%% mode: latex
%%% mode: visual-line
%%% fill-column: 5000
%%% TeX-master: "paper"
%%% End:

%  LocalWords:  omdoc-specific omdoc verb,isabelle mmt_build verb,mmt.jar verb,export_theory verb,export_standard_proofs primrec coinductive wrt consts Ballarin Ballarin2014 Nipkow-Prehofer:1993 fig:HOL-disj verb,id mathit verb,x verb,order verb,nat verb,less verb,less verb,c Longrightarrow sidewaysfigure HOL-disj.png verb,type verb,const verb,axiom verb,thm verb,ulo:uses verb,slow verb,large medskip

%% file: applications.tex
Our work now allows exporting entire Isabelle libraries into a format that can be easily read by third-party applications in a robustly maintainable way.
A major motivation for this work was enabling applications that use this exported data.
However, it remains open which applications should be better realized directly in Isabelle and which should be based on \mmt.
Critically, our export abstracts from most idiosyncrasies of Isabelle's logic, implementation, and library structure.
That has advantages and disadvantages.

On the positive side, any application that does not significantly depend on Isabelle's code base (e.g., search or dependency management) or explicitly rejects using it (e.g., representations in a logical framework or external proof checking) benefits from the uniform representation in the relatively simple language of \mmt.
On the negative side, any application that should be tightly integrated with Isabelle may be better realized natively in Isabelle.
This includes in particular applications that offer proof advice or rewrites/generates Isabelle data structures or Isabelle sources.

In some cases combined approaches may be indicated such as a small native addition to Isabelle that connects to a service implemented on top of the \mmt representation (and possibly running on a high-performance remote server).
For example, search services could be realized well in this way.
However, even when a native implementation that ignores the import into \mmt is indicated, our work can provide substantial benefits.
Any such native implementation will likely benefit from our streamlining and scaling up of Isabelle's export capabilities that allow integrating such applications with Isabelle.

Ultimately, the assessment which of these effects dominate must be made on a case-by-case basis for every application.
In the sequel, we sketch some applications enabled by our work where we expect the advantages to dominate.

\subsection{Clarification of Isabelle/Pure in Terms of MMT/PLF}

The Isabelle/Pure framework \cite{paulson700} is historically connected to Edinburgh LF, but it has its own distinctive style that can obscure important aspects.
The documentation \cite[\S2]{isabelle-implementation} refers to related formulations of $\lambda$\emph{HOL} within the setting of Pure Type Systems (PTS) due to Barendregt and Geuvers \cite{Barendregt-Geuvers:2001} and gives informal explanations (in {\LaTeX}) about how to understand Isabelle-specific concepts like schematic variables or type-classes.

Instead of Isabelle folklore and informal explanations in the documentation, our translation to PLF within \mmt elucidates many concepts of Pure more formally.
In particular:
\begin{compactitem}
\item The three levels of $\lambda$-calculus for function spaces (higher-order abstract syntax), universal binding of local parameters (quantification), logical entailment of rule statements (implication) become just one dependently-typed $\lambda$-calculus.
\item Implicit polymorphism becomes explicit as abstraction and quantification over types.
\item Up to scalability issues, proof terms --- which are an optional add-on to the Pure logic --- become plain $\lambda$-terms as definiens for theorems.
\item Type class constraints become explicit as predicates applied to types. Concretely, there are two possible representations for extra-logical constraints: \verb,'a::c, and intra-logical predication \verb|OFCLASS('a, c_class)|. Both are turned into the obvious term \verb,c a, for \verb,c :: type => prop, in PLF).
\end{compactitem}

Still lacking in our export is the explicit treatment of \emph{type class parameters}: as in Isabelle/Pure, the PLF theory treats instance-specific definitions as a collection of axioms that are associated with a generically typed constant.
A more sophisticated translation could try to make a dictionary construction, to turn type class parameters into explicit function parameters everywhere.

\subsection{External Proof Checking}

An often asked-for application of an Isabelle export is independent re-verification.
It may appear straightforward to use our export as the input of a separate application that specializes on re-checking proofs.
However, while this is certainly one of the intended uses, it would be naive to assume that our work is more than the first of multiple steps towards this goal.
In the sequel, we describe the remaining two obstacles: scalability and adequacy.
These obstacles are not inherent to our approach.
We expect any future solution to external proof checking to build on our approach or to recreate something comparable.

Regarding \textbf{scalability}, it is indeed straightforward to write a proof-checker for the Pure logic underlying Isabelle.
In fact, the \mmt formalization of Pure induces a proof-checker for Isabelle out of the box.
Similar framework-induced checkers can be built easily in implementations of LF-like frameworks such as Dedukti.
Moreover, the complexity of these checkers would typically be linear in the size of the proofs and thus very feasible.
It is even possible that checking the proofs could be faster than the file-system access needed to read the proofs in the first place.

But we do not expect such straightforward checkers to be able to handle the size of the proofs in the library:
the size of individual proofs, if naively encoded, may very well exceed the memory capacity of typical checkers.\footnote{Early experiments
  conducted with parts of the Main theory context of Isabelle/HOL produce hundreds of megabytes of proof terms in textual representation.}
Thus, additional investments are needed for handling large proofs, such as structure sharing, inferring omitted trivial steps, or streamed processing that can check a proof without loading it in its entirety.
These technologies are known in principle, but applying them to Isabelle/AFP remains substantial future work.

Regarding \textbf{adequacy}, note that our export is foundational in the sense that it exports the representation relative to the Pure logic in Isabelle's kernel, which arises from the original user input through a series of highly non-trivial transformations (elaboration).
Fully re-checking the proofs that result from elaboration is only one of two necessary conditions.
The other one is \emph{conservativity} of elaboration, i.e., the requirement that elaboration does not translate an unprovable statement to a provable one.
Depending on how many advanced Isabelle features are used in a problem statement, trusting the conservativity of elaboration may be a bigger leap than trusting the correctness of the proofs.

But conservativity is extremely difficult to establish.
The most direct way would be to specify the semantics of Isabelle's surface syntax and then prove Isabelle's elaboration algorithms correct relative to it.
Given the complexity of elaboration, this remains out of reach in the foreseeable future.

\subsection{Dependency Management}

The classic model of Isabelle/PIDE \cite{Wenzel:2014:ITP-PIDE} document markup merely provides a record of formal entities that are \emph{explicitly visible} in the source text.
Due to some reworking of the inference kernel by Wenzel, there is now a detailed record of all \verb,type, / \verb,const, / \verb,thm, entities that are \emph{implicitly used}.
This spans a rather large dependency graph over the original source: for Isabelle/AFP there are 400 million edges for 130\,MB of theory text.

In the past, users have occasionally attempted to approximate this information for their own purposes, e.g.\ in the Levity tool \cite{DBLP:conf/aisc/BourkeDKK12}, which exploits dependencies to move lemmas to adequate locations in the theory hierarchy.

Our ontological export (see Section~\ref{sec:ontology}) now includes a detailed record of both explicit source dependencies and implicit logical dependencies.
With this information available in a standard format, more ambitious (and more robust) refactoring tools can now be realized for Isabelle.
Optionally, such refactoring tools can even be built at the \omdoc/\mmt level in a way that they work uniformly for all systems that have exports similar to the one reported in this article.

\subsection{Search}

Because our export includes all logical information of the Isabelle content, it enables multiple search applications.
For example, this would allow searching for expressions or names that are not explicitly part of the sources and only occur in inferred information.
It also enables applying generic search systems to the Isabelle libraries.

As an example, we sketch a unification-based search services for the entire AFP based on MathWebSearch~\cite{KohSuc:asemf06}.
MathWebSearch maintains a substitution tree index that allows efficient unification queries over large collections of terms.
Because it can index \mmt terms, it can be directly applied to our export.
Thus, users can explore the full background library without having it loaded into the prover process (which might require too much memory), or even without installing the prover at all (e.g., by using a web service for the AFP).

Concretely, the queries would be terms with free variables over some AFP theory, and the search results would be terms in the AFP that unify with the query.
Because our export includes source references for all entities, these results can be linked to other resources (e.g., the location in the official AFP web site) or directly imported into PIDE.

The main remaining technical hurdle is the processing of the user's query.
In order to match anything in the library, formal objects in the query must be processed and exported in the same way as the library.
This includes the use of special forms for pattern matching, lists enumeration and comprehension etc. as well as type inference and type matching (with type classes).
Moreover, the user must provide the right context in which to interpret the query.

An intermediate solution could run a prover session of reasonable size that contains the most relevant notation (e.g., \verb,HOL-Analysis,) and process queries relative to it. These queries could then be exported and matched against the entire AFP.

We estimate that such a system is within reach of an ambitious Master's thesis.

%Further refinements can be envisioned for future work.
%For example, the Isabelle export already provides a full record of the algebra of type classes.
%A rather small Isabelle/Pure prover session could be used with this data to implement type inference and unification of the target context accurately, without loading the bulk of its theory content.
%This avoids re-implementing non-trivial prover operations outside of Isabelle/ML.

%A similar approach could work for namespace management, but the data for that is not exported yet (and more bulky).

\subsection{Enabling Cross-Library Knowledge Management}\label{sec:cross}

Isabelle/MMT is one of multiple large exports of proof assistant libraries that we have conducted over the last few years.
One of the original motivations of these efforts was to obtain multiple libraries in a uniform format in order to then develop develop cross-library and cross-prover knowledge management solutions.

These efforts are still at an experimental stage, and we only cite a few early results that could be extended to the Isabelle export:
\begin{compactitem}
\item We have used alignments \cite{KKMR:alignments:16} to relate corresponding concepts in different libraries.
  These can be annotated manually or found by machine learning techniques \cite{hol_isahol_matching}.
  Given a sufficient alignment coverage, we can then translate terms between libraries and use this to make systems interoperable.
\item With the relational RDF/XML export presented in Section~\ref{sec:ontology}, we can use SPARQL queries using the Upper Library Ontology (see~\cite{CKMRSW:ulo:19} for details) that return results from multiple libraries. \cite{BerKohRab:thqlmker20} presents an architecture for multi-aspect search based on these ideas. 
\item In~\cite{MKR:viewfinder:18} we have presented first steps towards finding views between different theorem prover libraries automatically.
%  This usually requires alignments for the logical languages of the libraries. \ednote{MK: say more about this, give examples. }
\end{compactitem}

%%% Local Variables:
%%% mode: latex
%%% mode: visual-line
%%% fill-column: 5000
%%% TeX-master: "paper"
%%% End:

%  LocalWords:  realizing paulson700 isabelle-implementation Geuvers Barendregt-Geuvers:2001 compactitem verb,c verb,c consts thms ednote conservativity KohSuc:asemf06 hol_isahol_matching KohMuePfe:kbimss17 omdoc BerKohRab:thqlmker20

%% file: conc.tex
\subparagraph{Summary}
In this article, we report on the conclusion of a research objective that seemed quite immediate two decades ago, but was not: the export of a theorem prover library (Isabelle) into a FAIR~\cite{WilDumAal:FAIR16} knowledge exchange format (\omdoc).
To make this undertaking feasible at all, both the source and target system had to evolve considerably:
Isabelle had to add its Scala and PIDE infrastructure to manage and expose document-oriented information in an instrumentable way, and the \omdoc format had to be re-engineered, extended, and implemented in the \mmt system.
Of course, the growth of the Isabelle library during this time induced further scalability problems, which we had to solve for our export.

Exports of theorem prover libraries have received substantial attention for the last 10--20 years.
Ours is the first comprehensive such export for Isabelle, and our work shows not only how to realize such an export for Isabelle but also the remaining theoretical and practical challenges.

Even ignoring the potential applications our particular export, our infrastructure for exporting Isabelle libraries in general will prove beneficial to future improvements to Isabelle itself and to the reuse of Isabelle content in other systems.
In fact, the improvements of Isabelle that were needed for our export have already shown benefits for the wider Isabelle community.
The \emph{headless PIDE} session and \verb,isabelle dump, tool have become particularly important: we are in personal contact with two different projects to build content-oriented search engines on top of these systems.
Another emerging application of this technology is a similar export of Isabelle to Dedukti \cite{dedukti}: this aims at re-checking the Isabelle/AFP and therefore includes proof terms but excludes PIDE document markup.

The current export facility is mostly based on code that is maintained within the Isabelle repository, and thus updated by the core developers.
We have already published Isabelle/\mmt for Isabelle2019 and Isabelle2020 based on a straight-forward process that users can easily recreate themselves: build \mmt within the Isabelle system environment, turn it into an Isabelle component, and use the standard Isabelle release tool to build a stand-alone variant of Isabelle that includes \mmt. %application (for Linux, Windows, and macOS).
Users can then rerun our export themselves on the spot (via the \verb,isabelle mmt_import, command).
We judge that this makes our Isabelle export the most easily reproducible and maintainable among all existing prover library exports.

\subparagraph{Future Work}
Besides realizing and scaling up the applications described in Section~\ref{sec:applications}, we want to mention two important avenues for future work:
\begin{compactitem}
\item The current export does not include proof objects as these would increase its size by an order of magnitude.
Instead, we restrict ourselves to the dependency relation induced by the proofs, which already enables many applications, but not, e.g., re-verification of proofs.  
To obtain scalable proof exports, we must investigate how to shrink the size of the proofs, e.g., by developing a new language for high-level proofs. 
\item In a similar vein we want to preserve the structure of more high-level declarations --- e.g. HOL-type definitions, inductive types.
As discussed in Section~\ref{sec:declarations}, this is supported by \mmt and would allow a structurally more similar and thus more understandable export. 
\end{compactitem}

%%% Local Variables:
%%% mode: latex
%%% mode: visual-line
%%% fill-column: 5000
%%% TeX-master: "paper"
%%% End:

%  LocalWords:  compactitem omdoc